\documentstyle[12pt]{article}
\begin{document}
\newcommand{\bfk}{{\mbox{\boldmath $k$}}}
\newcommand{\bfp}{{\mbox{\boldmath $p$}}}
\newcommand{\bfB}{{\mbox{\boldmath $B$}}}
\newcommand{\bfsig}{{\mbox{\boldmath $\sigma$}}}
\newcommand{\bfn}{{\mbox{\boldmath $n$}}}
\newcommand{\bfe}{{\mbox{\boldmath $e$}}}
\newcommand{\bfq}{{\mbox{\boldmath $q$}}}
\newcommand{\one}{{\mbox{1\hskip-0.5mm l}}}
\newcommand{\be}{\begin{eqnarray}}
\newcommand{\ee}{\end{eqnarray}}
\thispagestyle{empty}
\noindent
hep-ph/9510300
 \hfil \break
\rightline{PITHA 95/26}
\rightline{October, 1995}
\rightline{  }
\renewcommand{\thefootnote}{\fnsymbol{footnote}}
\begin{center}  {\Large\bf Transverse Polarization of Top
Quark Pairs\\ \vskip 3mm at the Tevatron and the Large Hadron
Collider{\footnote{supported by BMBF Contract 056AC92PE}}}
 \end{center}
\vskip 10 true mm
\begin{center} {Werner Bernreuther, Arnd Brandenburg, and Peter Uwer\\
\vskip 3mm
\it Institut f\"ur Theoretische Physik, Physikzentrum\\
Rheinisch-Westf\"alische Technische Hochschule Aachen\\
52056 Aachen, Germany} \end{center}
\vskip 10 true mm

\bigskip
\centerline{\large\bf Abstract}
\smallskip
\noindent  We investigate the prospects to observe effects
of transverse
polarization of top quarks in $pp,p\bar{p}\to t\bar{t}X$.
QCD absorptive parts generate a polarization of top
quarks and antiquarks transverse to the production plane in the partonic
processes $q\bar{q}\to t\bar{t}$ and $gg\to t\bar{t}$, which reaches
values of a few percent. These perturbative QCD effects are decreased at
the hadronic level. A measurement through
momentum correlations among the $t$ and $\bar{t}$ decay
products
will be difficult both at an upgraded
Tevatron and at the LHC.
\vfil\eject
\setcounter{page}{1}
\bigskip
\section{Introduction}
The existence of the top quark has recently been firmly established
\cite{CDFD0}. Combining the measurements of the CDF and D0 experiments,
the mass
of the top quark is $m_t=179\pm 12$ GeV. Detailed studies of the properties
of top quarks will be a main objective of experiments at present and future
colliders. With an expected luminosity  of ${\cal L}\sim 2\times 10^{32}
{\mbox{cm}}^{-2}{\mbox s}^{-1}$, the upgraded Tevatron
collider will produce several
thousand top quark pairs per year, and the Large Hadron Collider (LHC) is even
expected to yield several million top quark pairs per year. These
event rates will
allow for precision experiments with top quarks. \par
A special feature of physics with top quarks is due to their
heaviness: Because of its large mass, the top quark  has decayed
on average before it can form hadronic bound states. In particular, the
polarization of the top quark is not diluted by hadronization, and observables
involving the spin of the top quark may therefore be calculated reliably
within perturbation theory. Moreover, the $t$ and $\bar{t}$ analyze their
spins through their parity-violating weak decays $t\to Wb$.
Effects connected with the spin of the top quark
can then be used to test the standard model (SM) or to search for new physics,
 as has been realized and investigated by many authors (see, e.g.,
\cite{spinreffirst}--\cite{spinreflast}, and references quoted
therein).\par
In this paper we will concentrate on studying a possible polarization
of the $t$ and $\bar{t}$ transverse to the production plane in proton-proton
and proton-antiproton collisions. This transverse polarization is odd
under simultaneous reflection of spins and momenta (i.e., odd under
the ``naive''
time reversal operation $T_N$).  To leading order, it is directly
proportional to the
absorptive part of the scattering amplitude if the true time reversal
operation is a good symmetry of the theory. In particular, it probes QCD
at the one-loop level without having to  contend with the usual
large tree level
``background''.\par
The main production mechanism for $t\bar{t}$-pairs in hadronic collisions are
the parton subprocesses $q\bar{q}\to t\bar{t}$ and $gg\to t\bar{t}$. QCD
absorptive parts giving rise to a transverse polarization of the top quark
and antiquark have been calculated for these parton processes first by
Dharmaratna and Goldstein \cite{Goldstein1},\ \cite{Goldstein2}. Their
result for $gg\to t\bar{t}$ was used by Kane, Ladinsky and Yuan
to estimate the size of transverse
polarization to be expected at LHC energies and beyond \cite{KLY}.
We reanalyze
this subject for two reasons: First, because the process
$q\bar{q}\to t\bar{t}$ is dominant at the Tevatron and the corresponding
transverse
polarization phenomena at the hadronic level have not been studied before.
Second, we would like to present realistic numbers for directly measurable
quantities sensitive to the transversely polarized top quarks. These
quantities are $T_N$--odd momentum correlations of the decay products
of the $t$ and $\bar{t}$. The outline of our paper is as follows:
In the next section, we will give analytic results for the transverse
polarization of $t$ and $\bar{t}$ in
$q\bar{q}\to t\bar{t}$ and $gg\to t\bar{t}$
to order $\alpha_s$, point out a slight difference
to existing results \cite{Goldstein1}, and discuss the effects on parton
level. In section 3, we will proceed by constructing suitable observables
to trace these effects in top pair production at hadron
colliders. The corresponding $T_N$--odd momentum
correlations will require a measurement
of momenta of final
state particles only. The statistical significance of these correlations
will be discussed for the Tevatron and the LHC.

\section{Transverse Polarization in $q\bar{q}\to t\bar{t}$ and
$gg\to t\bar{t}$}
In this section we will define transverse polarization of top quarks and
antiquarks in terms of the production density matrices for the partonic
subprocesses $q\bar{q}\to t\bar{t}$ and $gg\to t\bar{t}$, which are the
dominant production mechanisms for top quark pair production at the
Tevatron and the LHC, respectively. We will give analytic results for
the transverse polarization of $t$ and $\bar{t}$ induced by QCD absorptive
parts at the one-loop level.\par
We first turn to the reaction
$q(p_1)+\bar{q}(p_2)\to t(k_1)+\bar{t}(k_2)$, where the momenta refer to the
$q\bar{q}$ center of mass system, $\bfp_1+\bfp_2=0$. In order to discuss
polarization phenomena for $t$ and $\bar{t}$, we define the (unnormalized)
production density matrix
\begin{eqnarray}
R^q_{\alpha_1\alpha_2,\beta_1\beta_2}(\bfp,\bfk)=
\frac{1}{4}\frac{1}{N_C^2}\sum_{{\mbox{\scriptsize{colors}}},
q\bar{q}{\mbox{\scriptsize{\ spins}}}}
\langle t(k_1,\alpha_1)\bar{t}(k_2,\beta_1)|{\cal{T}}|
q(p_1),\bar{q}(p_2)\rangle^* & &\nonumber\\
\langle t(k_1,\alpha_2)
\bar{t}(k_2,\beta_2)|{\cal{T}}|
q(p_1),\bar{q}(p_2)\rangle. & &
\end{eqnarray}
Here, $\alpha,\ \beta$ are spin indices, $N_C$ denotes
the number of colors, $\bfp=\bfp_1,\ \bfk=\bfk_1$ and
the sum runs over the colors of all quarks and over the spins of $q$ and
$\bar{q}$. The factor $1/4\cdot 1/N_C^2$ takes care of the averaging over
spins and colors in the initial state. The matrix structure
of $R^q$ in the spin spaces of $t$ and $\bar{t}$ is
\begin{eqnarray}
R^q=A^q\one\otimes \one+ \bfB^q_t\cdot\bfsig\otimes \one+\bfB^q_{\bar{t}}\cdot
\one\otimes\bfsig+C^q_{ij}\sigma^i\otimes\sigma^j,
\end{eqnarray}
where $\sigma^i$ are the Pauli matrices and the first (second) factor in the
tensor products refers to the $t$ ($\bar{t}$) spin space. For a detailed
discussion of $R^q$ and its symmetry
properties, see \cite{Bernbra}. For the purpose of this paper,
it is sufficient to note that absorptive parts
in the parton scattering amplitude due to $CP$- and $P$- invariant
interactions yield the following contribution to the production density
matrix:
\begin{eqnarray}
R_{{\mbox{\scriptsize{abs}}}}^q=b_3^q(\hat{s},z)\hat\bfn\cdot
\left\{\bfsig\otimes \one+\one\otimes\bfsig\right\},
\end{eqnarray}
where $\hat{s}=(p_1+p_2)^2$, $z=\hat\bfp\cdot\hat\bfk$ is the cosine
of the scattering
angle in the partonic center of mass frame,
$\hat\bfn=(\bfp\times\bfk)/|\bfp\times\bfk|$ is the normal to the
scattering plane, and $b_3^q(\hat{s},z)$ is a structure function in
the notation of \cite{Bernbra}.
The production density matrix $R^g(\bfp,\bfk)$ for the
subprocess $g(p_1)+g(p_2)\to t(k_1)+\bar{t}(k_2)$ is defined, in the
$gg$ center of mass frame, in complete analogy to Eq. (1) with
$1/N_C^2\to 1/(N_C^2-1)^2$. Due to Bose symmetry of the
gluon-gluon initial state we have for the corresponding structure
function:
\begin{eqnarray}
b_3^g(\hat{s},z)=-b_3^g(\hat{s},-z).
\end{eqnarray}
The transverse polarization $P_{\bot}$ of the top quark is equal to the
transverse polarization of the top antiquark in a CP invariant theory
and given for the respective subprocesses by
\begin{eqnarray}
P_{\bot}^{i}=\langle \hat\bfn\cdot\bfsig\otimes\one\rangle_{i}
\equiv\frac{{\mbox{tr}}(R^{i}
\ \hat\bfn\cdot\bfsig\otimes\one)}{{\mbox{tr}}
\ R^{i}}=\frac{b_3^{i}(\hat{s},z)}{A^{i}(\hat{s},z)}\ \ \ \ \ \ (i=q,g).
\end{eqnarray}
The functions $A^{i}(\hat{s},z)$ are related to the differential
cross section
of the respective partonic subprocess:
\be
A^{i}(\hat{s},z)=\frac{8\pi\hat{s}}{\beta}\frac{d\hat{\sigma}^{i}}{dz},
\ee
where $\beta=\sqrt{1-4m_t^2/\hat{s}}$ is the velocity of the top quark
in the partonic c.m. system.
In leading order of the QCD coupling,
they have the well-known form \cite{GORC}
(which we give here for completeness):
\be
A^q(\hat{s},z)\ &=&\ \pi^2\alpha_s^2\frac{N_C^2-1}{N_C^2}
\left[2-\beta^2(1-z^2)\right],\\
A^g(\hat{s},z)\ &=&\  \pi^2\alpha_s^2\frac{2\left[-2+N_C^2(1+\beta^2z^2)
\right]}
{N_C(N_C^2-1)(1-\beta^2z^2)^2}\left[1+2\beta^2(1-z^2)(1-\beta^2)-\beta^4z^4
\right].
\ee
In $A^q$ we neglected the masses of the quarks in the initial state.
\par
Nonvanishing $b_3^{i}(\hat{s},z)$ are generated by QCD absorptive parts at
the one-loop level through interference with the tree graphs.
Other standard model contributions, like from absorptive parts
of diagrams with Higgs or $Z$ bosons, are tiny and therefore we
neglect them.
Before presenting our results, we would like to comment briefly
on the details of the calculation.
We performed
two independent calculations of the functions  $b_3^{i}(\hat{s},z)$,
 one using conventional covariant gauge
Feynman rules, the other using the background field method \cite{Abbott}.
The background field method greatly facilitates the calculation
of $b_3^g$, and, because of gauge invariance, gives the same final
result as the conventional approach.
In Fig. 1 (2) we show the one-loop Feynman diagrams which give
a nonzero contribution to $b_3^q$ ($b_3^g$) in the conventional
approach. Although
the final results are ultraviolet and infrared finite, one has to use a
regularization scheme in intermediate steps to handle divergent integrals.
We used dimensional regularization. Moreover, we eliminated the usual
box integral in four dimensions in favor of a linear combination
of a six-dimensional box integral and four triangle integrals \cite{Dixon}.
This methods shifts all infrared divergencies into the triangle integrals,
and the cancellation of the infrared divergencies becomes completely
explicit.
\par
For $q\bar{q}\to t\bar{t}$ we get:
\begin{eqnarray}
b_3^q\ &=&\  \frac{\pi \alpha_s^3(N_C^2-1)}{2N_C^2}
\frac{m_t}{\sqrt{\hat{s}}}\sqrt{1-z^2}
\big(N_C g_1+\frac{g_2}{N_C}\big),\nonumber \\  {\mbox{with}}
\ & &\  \nonumber \\
g_1 &=& -\frac{\pi}{\beta^2}\bigg[(1-\beta^2)
\ln\left(\frac{1+\beta}{1-\beta}\right)-2\beta\bigg]
-\frac{\pi z}{2\beta^3}\bigg[3(1-\beta^2)\ln\left
(\frac{1+\beta}{1-\beta}\right)-6\beta+4\beta^3\bigg], \nonumber \\
g_2 &=& \frac{4\pi}{\beta^2}\bigg[ (1-\beta^2)\ln\left(\frac{1+\beta}
{1-\beta}
\right)
-2\beta\bigg]+\pi\beta z. \label{qqres}
\end{eqnarray}
This result agrees with the one given in \cite{Goldstein2}.\par
For $gg\to t\bar{t}$, the result is more complex:
\begin{eqnarray}
b_3^g &=& \frac{2\pi \alpha_s^3}{N_C^2-1}m_t\beta \sqrt{\hat{s}}
\sqrt{1-z^2}
\big(N_C^2 f_1+N_C f_2+f_3  +\frac{f_4}{N_C^2}\big),\nonumber \\
{\mbox{with}} \ & &\
\nonumber \\
f_1 &=& \frac{1-\beta^2}{2(1-z\beta)^2}{\mbox{Im}} D_0^{D=6}(z)
         -\frac{1-\beta^2}{2(1+z\beta)^2}{\mbox{Im}} D_0^{D=6}(-z)
         \nonumber \\ \ & &\
         -\frac{z(1-\beta^2)(1-8\beta^2-3z^2-\beta^2z^2+3\beta^2z^4)}
               {4\beta(1-z^2\beta^2)^2}{\mbox{Im}}C_0
          \nonumber \\ \ & &\
         -\frac{z(1-4\beta^2-3z^2+2\beta^2z^2)}{2\beta \hat{s}
(1-z^2\beta^2)
               }{\mbox{Im}}B_0, \nonumber \\
f_2\ &=&\ \sum_i\theta(\sqrt{\hat{s}}/2-m_i)\frac{z\beta(1-\beta_i^2)}
          {2(1-z^2\beta^2)}
         \left[\frac{\pi}{\hat{s}}\ln\left(\frac{1+\beta_i}
         {1-\beta_i}\right)-\frac{2\pi\beta_i}{\hat{s}}\right], \nonumber
\\
f_3\ &=&\ \frac{\beta}{4(1-z^2\beta^2)}
\bigg[
-(3\beta^2+z^2-6\beta z+2\beta^2z^2){\mbox{Im}}{D}_0^{D=6}(z)
\nonumber \\ \ & & \ +
(3\beta^2+z^2+6\beta z+2\beta^2z^2){\mbox{Im}}{D}_0^{D=6}(-z)
-\frac{2z(1-\beta^2)}{\beta}{\mbox{Im}}C_0
\nonumber\\ \ & &\  -
\frac{2z(2+5\beta^2-\beta^2z^2)}
{\hat{s}\beta}{\mbox{Im}}B_0\bigg], \nonumber \\
f_4\ &=&\ \frac{\beta}{2(1-z^2\beta^2)}
\bigg[
-\frac{(3\beta^2+z^2-6\beta z+2\beta^2z^2)}{1-z\beta}
{\mbox{Im}}{D}_0^{D=6}(z)\nonumber \\ \ & &\ +
\frac{(3\beta^2+z^2+6\beta z+2\beta^2z^2)}{1+z\beta}
{\mbox{Im}}{D}_0^{D=6}(-z)
\nonumber\\ \ & &\
-\frac{2z(1-\beta^2)(1-3\beta^2)}{\beta(1-z^2\beta^2)}
{\mbox{Im}}C_0
-\frac{4z}{\hat{s}\beta}{\mbox{Im}}B_0\bigg]. \label{ggres}
\end{eqnarray}
Here, $D_0^{D=6}$ stands for a box integral in six dimensions, $B_0$
and $C_0$ are usual two-point and three-point scalar
one-loop integrals, respectively. The explicit formulae for
the imaginary parts are:
\begin{eqnarray}
 {\mbox{Im}}D_0^{D=6}(z)\ &=&\
-\frac{\pi}{\hat{s}\beta^2}
\left\{\frac{1+\beta}{1+z}\ln\left(1-\frac{\beta(1+z)}{1+\beta}
\right)+\frac{1-\beta}{1-z}
\ln\left(1+\frac{\beta(1-z)}{1-\beta}\right)\right\}, \nonumber \\
{\mbox{Im}}C_0 \ &=&\
\frac{-\pi}{\hat{s}\beta}\ln\left(\frac{1+\beta}{1-\beta}\right), \nonumber \\
{\mbox{Im}}B_0\ &=&\ \pi.
\end{eqnarray}
The sum in $f_2$ defined in Eq. (\ref{ggres}) runs over all quark flavors and
$\beta_i=\sqrt{1-4m_i^2/\hat{s}}$, where $m_i$ denotes
the mass of a quark with flavor $i$. Numerically, only the top quark
gives a significant contribution to $f_2$.
Our result for $b_3^g$
disagrees slightly with the one given in formula (1)
of \cite{Goldstein1}
(and formula (19) of \cite{Goldstein2}).
We find that
 $\displaystyle f_1-f_1^{DG}=
-\frac{\pi z\beta}{48\hat{s}(1-z^2\beta^2)}
$, where $f_1$ is the leading color coefficient of Eq. (\ref{ggres})
and $f_1^{DG}$ is the corresponding term in
\cite{Goldstein1}, \cite{Goldstein2}. We agree{\footnote{G. Goldstein,
private communication}} with the authors of
\cite{Goldstein1}, \cite{Goldstein2} that this difference is due to
the omission of one of the two ghost triangle diagrams (cf. Fig. 2) in the
result of \cite{Goldstein1}, \cite{Goldstein2}.
\par
In Figs. 3 and 4 we plot the transverse polarization $P_{\bot}^q$ and
$P_{\bot}^g$, respectively, as a function of the partonic c.m. energy
$\sqrt{\hat{s}}$ and $z$, the cosine  of the scattering angle.
We do not include the running of $\alpha_s$ here, but
fix it at the value $\alpha_s=0.1$. For the top quark mass
we used $m_t=180$ GeV. For quark--antiquark
annihilation, the transverse polarization of the top quark
reaches values of about $2.5\%$ around $\sqrt{\hat{s}}\simeq 720$ GeV
and scattering angle of $\simeq 73$ degrees  and then
decreases quite
rapidly with energy. In the case of gluon--gluon fusion, $P_{\bot}^g$ nicely
exhibits the antisymmetry with respect to $z$ following from Bose symmetry
and  $P_{\bot}^g$ reaches peak values of about $1.5\%$
around $\sqrt{\hat{s}}\simeq 1050$ GeV and at $\simeq \pm 63$ degrees.
We will discuss in the next section whether these effects are observable
at colliders.

\section{$T_N$--odd momentum correlations for
semileptonic $t\bar{t}$ decays}

The transverse polarization of $t$ and $\bar{t}$ discussed in the previous
section must be traced in the final states into which $t$ and
$\bar{t}$ decay. We concentrate here on decay modes where one of the
top quarks decays semileptonically and the other one decays hadronically,
i.e. on samples of $t\bar{t}$ pairs with:
\be
& & t\to \ell^++\nu_{\ell}+b,\nonumber \\
& & \bar{t} \to W^-+\bar{b}\to q+\bar{q}'+\bar{b},\label{sample}
\ee
and the corresponding charge conjugated decay channels.
These samples are especially suited for constructing observables
which are sensitive to transversely polarized top quarks: The charged
lepton of the semileptonic decay serves as an efficient analyzer
of the top quark spin \cite{Czar},
while in the same event the purely hadronic decay
allows one to reconstruct the momentum of the other top quark.
Specifically, for $p\bar{p}\to t\bar{t}X$
with subsequent semileptonic
$t\bar{t}$ decay we define the observable
\be
{\cal O}_1= \hat\bfp_p\cdot(\hat{\bfq}_{{\ell}^+}
\times \hat{\bfk}_{\bar{t}}) \label{Op}
\ee
for the decay modes (\ref{sample}), and
\be
\bar{\cal O}_1= \hat\bfp_p\cdot(
\hat{\bfq}_{{\ell}^-}\times \hat{\bfk}_{t}) \label{Om}
\ee
for the charge conjugated decays. Here, $\hat\bfp_p$ is the
proton's direction (which we take to be the positive z direction of the
laboratory coordinate system), $\bfq_{\ell^{\pm}}$ is the momentum
of the positively (negatively) charged lepton,
hats denote unit vectors  and all momenta
are defined in the hadronic c.m. system.
A $T_N$--odd correlation may now be defined through the sum:
\be
{\cal S}_1\equiv\langle{\cal O}_1\rangle_{\bar{t}\ell^{+}}+
         \langle\bar{\cal O}_1\rangle_{t\ell^{-}}, \label{ppbarobs}
\ee
where  $(\bar{t}\ell^+)$ refers to sample (\ref{sample})
(reconstructed flight directions of $\bar{t}$ and of the positively charged
lepton from $t$ decay) and $(t\ell^-)$ refers to the charge conjugated sample.
Gluon--gluon fusion does not contribute
to either of the two terms on the r.h.s. of (\ref{ppbarobs}) due
to Bose symmetry, but this does not lead to a significant decrease
of the signal in the case of $p\bar{p}$ collisions around a c.m. energy
$\sqrt{s}=1.8$ TeV, because in this energy regime the contribution
of the partonic subprocess $q\bar{q}\to t\bar{t}$ strongly dominates.
Since the unpolarized $p\bar{p}$ initial state
is a $CP$ eigenstate, the correlation
(\ref{ppbarobs}) can only
become nonzero through $CP$ conserving interactions.
Furthermore, $\langle{\cal O}_1\rangle_{\bar{t}\ell^{+}}=
\langle\bar{\cal O}_1\rangle_{t\ell^{-}}$, if $CP$ is conserved. Taking the
difference instead of the sum in (\ref{ppbarobs}) gives
a $T_N$-odd, $CP$-odd correlation, which might be also an interesting
quantity to look at in $t\bar{t}$ production.\par
For $pp$ collisions, each of the two expectation
values in (\ref{ppbarobs}) vanishes
for the following reason: The contribution
from $gg\to t\bar{t}$ again vanishes
due to Bose symmetry, and also the subprocess $q\bar{q}\to t\bar{t}$
now gives zero, because the contributions of the two
partonic initial states $(q\in p(\bfp_p),\bar{q}\in p(-\bfp_p))$ and
$(\bar{q}\in p(\bfp_p),q\in p(-\bfp_p))$ cancel. Suitable observables
in the case of $pp$ collisions are given by:
\be
{\cal O}_2= -{\cal O}_1 \ {\mbox{sign}}(\hat\bfp_p\cdot\hat\bfk_{\bar{t}}),
\label{Opfb}\ee
for sample (\ref{sample}), and
\be
\bar{\cal O}_2= \bar{\cal O}_1
\ {\mbox{sign}}(\hat\bfp_p\cdot\hat\bfk_{t})
\label{Omfb}\ee
for the charge conjugated sample.
The sum of the expectation values,
\be {\cal S}_2\equiv\langle{\cal O}_2\rangle_{\bar{t}\ell^{+}}+
         \langle\bar{\cal O}_2\rangle_{t\ell^{-}}, \label{ppobs}
\ee
is a good $T_N$--odd correlation to trace the transverse polarizations
of $t$ and $\bar{t}$ produced in $pp$ collisions.
Note that $\langle{\cal O}_2\rangle_{\bar{t}\ell^{+}}\not=
\langle\bar{\cal O}_2\rangle_{t\ell^{-}}$ even in the absence
of $CP$ violation.
This is because the $pp$ initial state has no definite
$CP$ parity.
In fact,
since quarks on average carry more of the proton's energy
than antiquarks, QCD absorptive parts in the $q\bar{q}$
annihilation subprocess lead to a nonvanishing value for
the difference
\be {\Delta}_2\equiv\langle{\cal O}_2\rangle_{\bar{t}\ell^{+}}-
         \langle\bar{\cal O}_2\rangle_{t\ell^{-}} \label{ppobsplus}
\ee
in $pp$ collisions.\par
We evaluated the correlation ${\cal S}_1$ for $p\bar{p}$ c.m.
energies $\sqrt{s}$ between 1.5 TeV and 5.0 TeV and the two
correlations  ${\cal S}_2$ and
 $\Delta_2$ for $pp$ collisions with $\sqrt{s}$ between
7 TeV and 20 TeV,
using the narrow width approximation for $t$ and $\bar{t}$.
(For a recent discussion of effects of non-factorizable
diagrams see \cite{Melnikov}.)
We found only a weak dependence of our results
on the choice of the parton distribution functions; in the results below
we  used the
parametrization \cite{DO}. For the scale entering the parton distributions
we set $Q^2=4m_t^2$. Since the results of sect. 2 constitute
the lowest
order contributions to our correlations, we  use leading order
distribution functions and a fixed  value of $\alpha_s=0.1$.\par
Fig. 5 shows the correlation ${\cal S}_1$ for proton--antiproton
scattering as a function of the collider c.m. energy.
The correlation
decreases with rising energy; at $\sqrt{s}=1.8$ TeV it has the value
${\cal S}_1\simeq 0.43$\%.
The rapidity distribution
\be \langle{\cal O}_1\delta(r_{\bar{t}}-r'_{\bar{t}})\rangle_
{\bar{t}\ell^{+}}\label{raptbar}\ee
where
\be r_{\bar{t}}=\frac{1}{2}\ln\frac{E_{\bar{t}}+\hat\bfp_p\cdot \bfk_{\bar{t}}}
{E_{\bar{t}}-\hat\bfp_p\cdot \bfk_{\bar{t}}} \ee
is shown in Fig. 6
for $\sqrt{s}=1.8$ TeV. Note that the distribution
is not symmetric in $r_{\bar{t}}$. The
corresponding rapidity distribution for the
sample with
reconstructed top momenta,
 \be\langle\bar{\cal O}_1\delta(r_t-r'_t)\rangle_{t\ell^{-}}
\label{rapt}\ee
is equal to the one of Fig. 6 after a reflection at the line
$r_{\bar{t}}=0$.\par
Fig. 7 shows both ${\cal S}_2$ and  $\Delta_2$ for
proton--proton collisions with c.m. energies covering 7--20 TeV.
The correlation ${\cal S}_2$ reaches a value of $\sim 0.03$ \% at
$\sqrt{s}=14$ TeV.
Fig. 8 shows the rapidity distributions (\ref{raptbar})
and (\ref{rapt}) for $pp$ collisions at $\sqrt{s}=14$ TeV. As discussed
above, they are {\it {not}} related by a symmetry.\par
The statistical sensitivity of our observables may be estimated
from their root-mean-square fluctuations: For
$\sqrt{s}=1.8$ TeV we find $\Delta {\cal O}_1=
\sqrt{\langle{\cal O}_1^2\rangle
_{\bar{t}\ell^{+}}
-\langle{\cal O}_1\rangle_{\bar{t}\ell^{+}}^2}\approx\Delta
\bar{\cal O}_1\simeq 0.36$.
This yields an estimate for the 1 s. d. statistical
error $\delta {\cal S}_1\simeq (\sqrt{2}\times 0.36)/
\sqrt{N_{event}}\simeq 0.5/\sqrt{N_{event}}$,
where $N_{event}$ is the number of events of type (\ref{sample}).
In order to push $\delta {\cal S}_1$ below the
percent level $N_{event}> 10^4$ events
are needed  -- an unrealistic
number even for the upgraded
Tevatron. In the case of $pp$ collisions at the LHC energy
$\sqrt{s}=14$ TeV, we have $\Delta {\cal O}_2\approx\Delta
\bar{\cal O}_2 \simeq 0.25$, which implies
$|{\cal S}_2/(\sqrt{2}\Delta {\cal O}_2)|\simeq 10^{-3}$.
Although the production of
$10^7$ $t\bar{t}$ pairs may be expected
at the Large Hadron Collider operating at high luminosity, the systematic
errors of a measurement of our correlations
at a hadron collider will probably
be too large to probe effects of the order of a per mill.\par
One may ask whether there are better observables to trace
transverse polarization than the ones we chose. Obvious variations are,
for example, $\hat\bfp_p\cdot (\bfk_{\bar{t}}\times
\bfq_{\ell
^{+}})/s$ or $\hat\bfp_p\cdot (\hat\bfk_{\bar{t}}\times \bfq_{\ell
^{+}})/\sqrt{s}$. Another variation would be
$\hat\bfp_p\cdot (\bfk_{\bar{t}}\times \bfq_{\ell
^{+}})/|(\bfk_{\bar{t}}\times \bfq_{\ell
^{+}}|$. All of these observables lead to similar or smaller ratios
${\cal O}/\Delta {\cal O}$. We did not attempt to construct fully
optimized observables
\cite{AtNacht} here. A more thorough analysis would construct such
observables, which maximize the signal-to-noise ratio, not only for
the semileptonic $t\bar{t}$ decay channels discussed in this
paper, but also
for all other decay modes \cite{bu}.\par
In conclusion we have computed the QCD-induced transverse polarization
of $t$ and $\bar{t}$ quarks produced in high energetic $p\bar{p}$ and
$pp$ collisions. Resulting angular correlations which involve the
$t$ or $\bar{t}$ decay products are at the percent level at Tevatron
energies and at the per mill level at the LHC. As these are definite QCD
predictions it will still be worthwhile to measure the correlations
(\ref{ppbarobs}), (\ref{ppobs}), since they can be used as ``sensors''
for new physics effects in the top system: any observable effect --- i.e.,
any significant deviation from the above results --- would point
towards strong $t\bar{t}$ final state interactions beyond the Standard
Model forces.

\bigskip
\noindent {\bf Acknowledgements}\par\noindent
A. B. would like to thank L. Dixon and Y. Shadmi for useful discussions.

\vfil\eject
{\Large{\bf Figure Captions}}
\medskip
\begin{description}
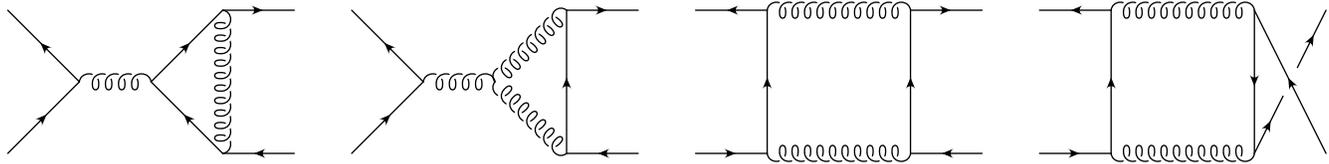
\item[Fig. 1] One-loop Feynman diagrams for the process
 $q\bar{q}\to t\bar{t}$ which give a nonzero con\-tribution
to the function $b_3^q$ defined in Eq. (3) through
interference with the Born diagram.
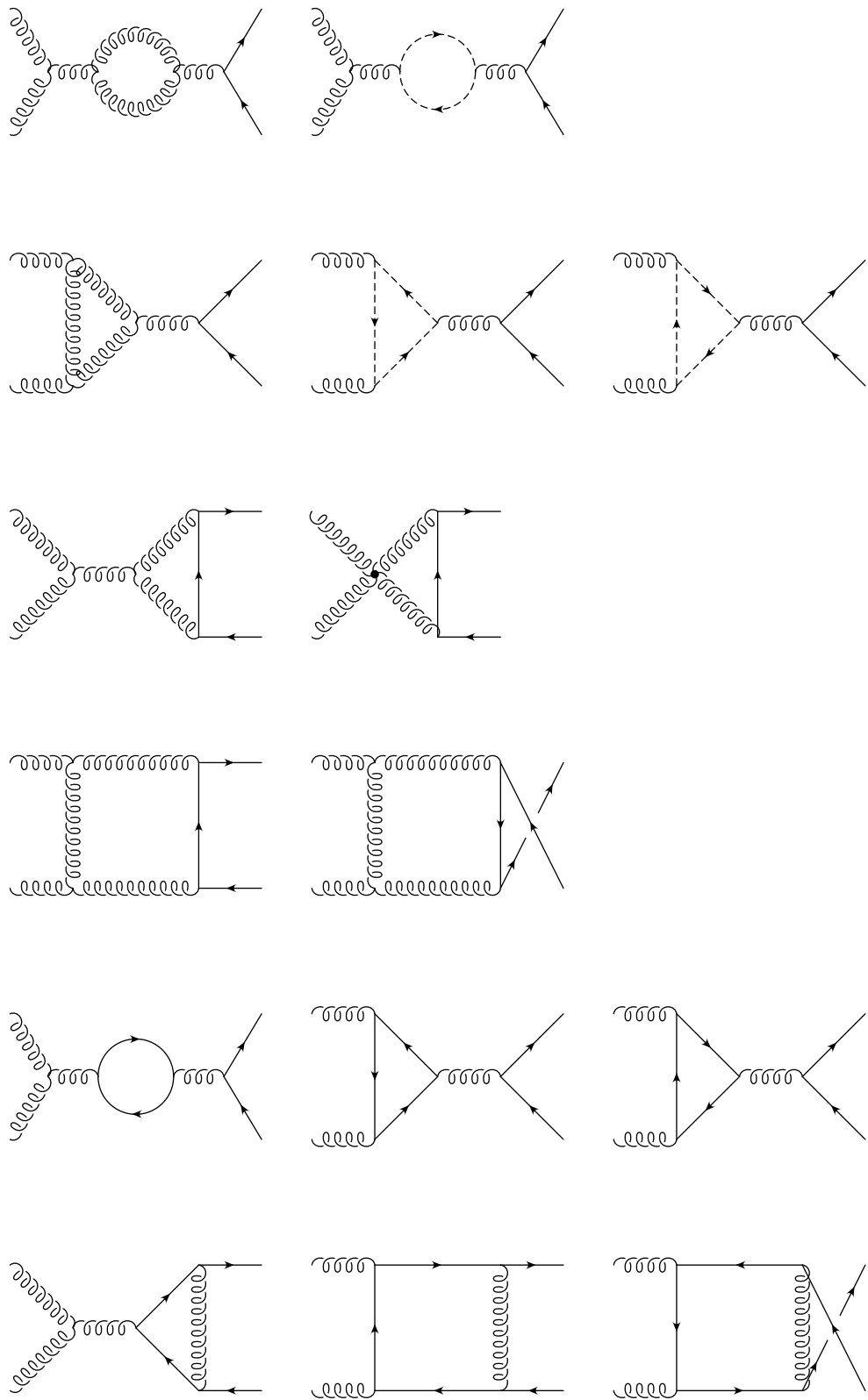
\item[Fig. 2] One-loop Feynman diagrams for the process
 $gg\to t\bar{t}$ which give a nonzero con\-tribution
to the function $b_3^g$ through interference with the Born
diagrams.
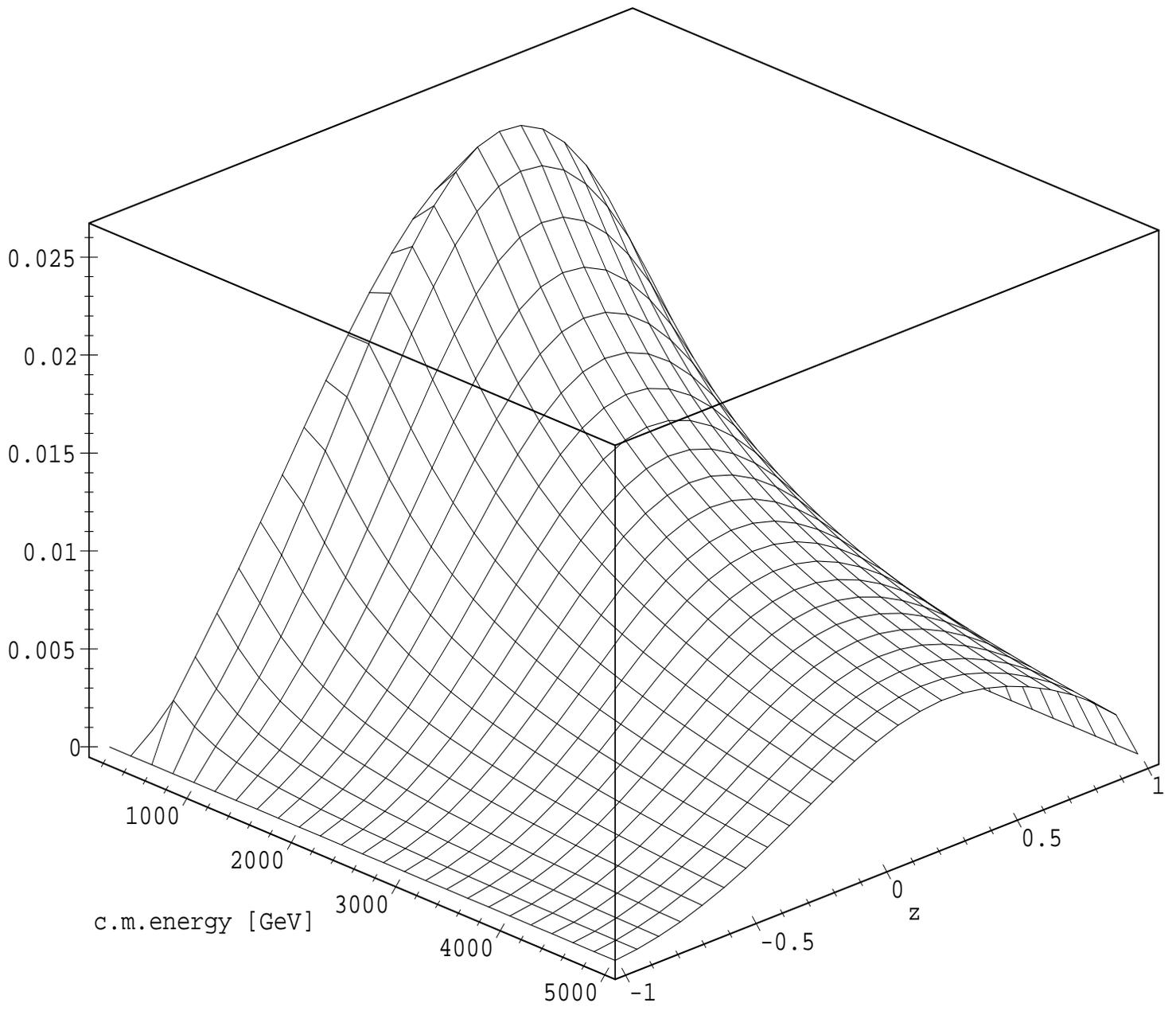
\item[Fig. 3] Transverse polarization $P_{\bot}^q$ (defined
in Eq. (5)) as a function of the partonic c.m. energy $\sqrt{\hat{s}}$
and the cosine of the scattering angle $z$.
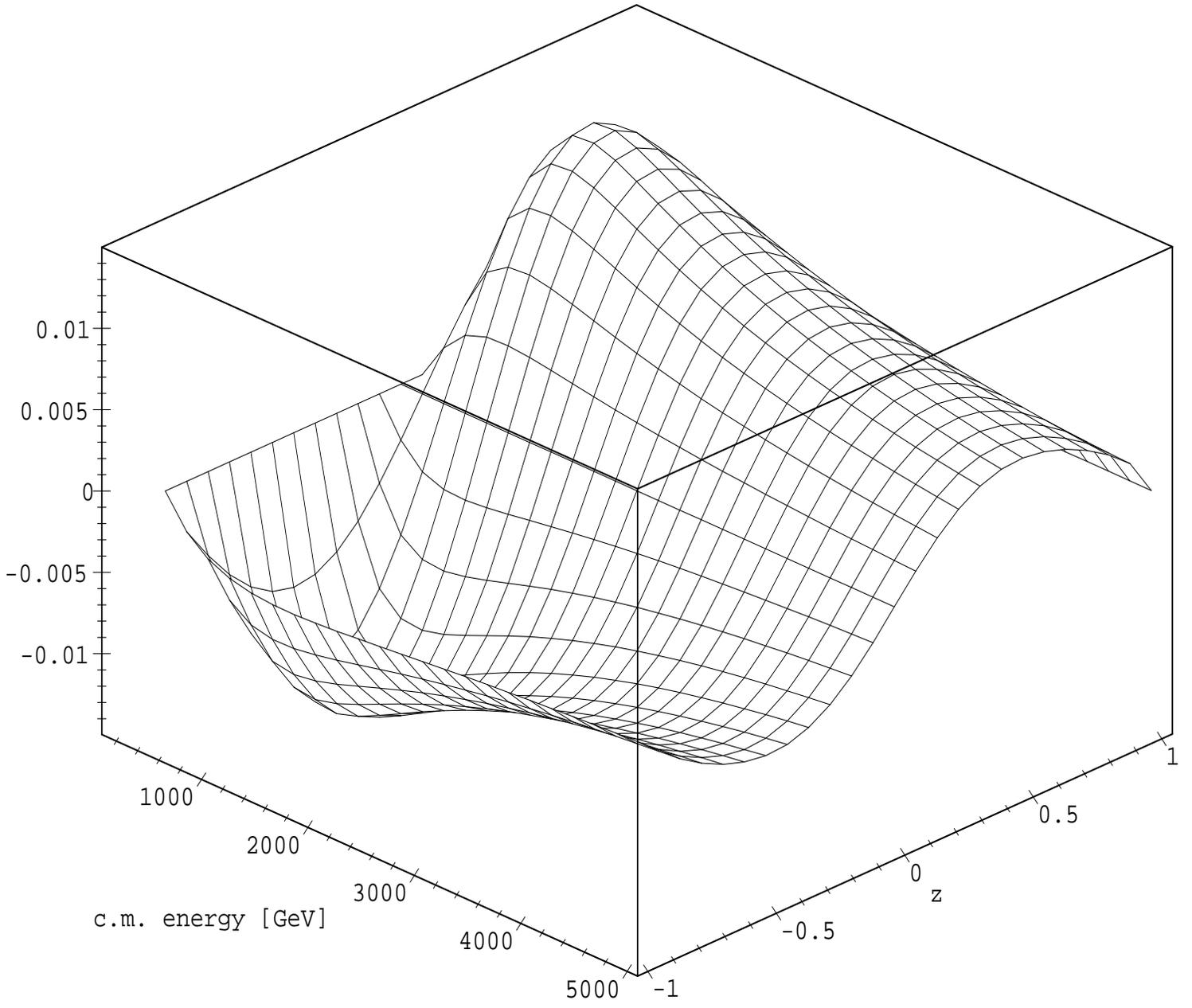
\item[Fig. 4] Transverse polarization $P_{\bot}^g$ (defined
in Eq. (5)) as a function of the partonic c.m. energy $\sqrt{\hat{s}}$
and the cosine of the scattering angle $z$.
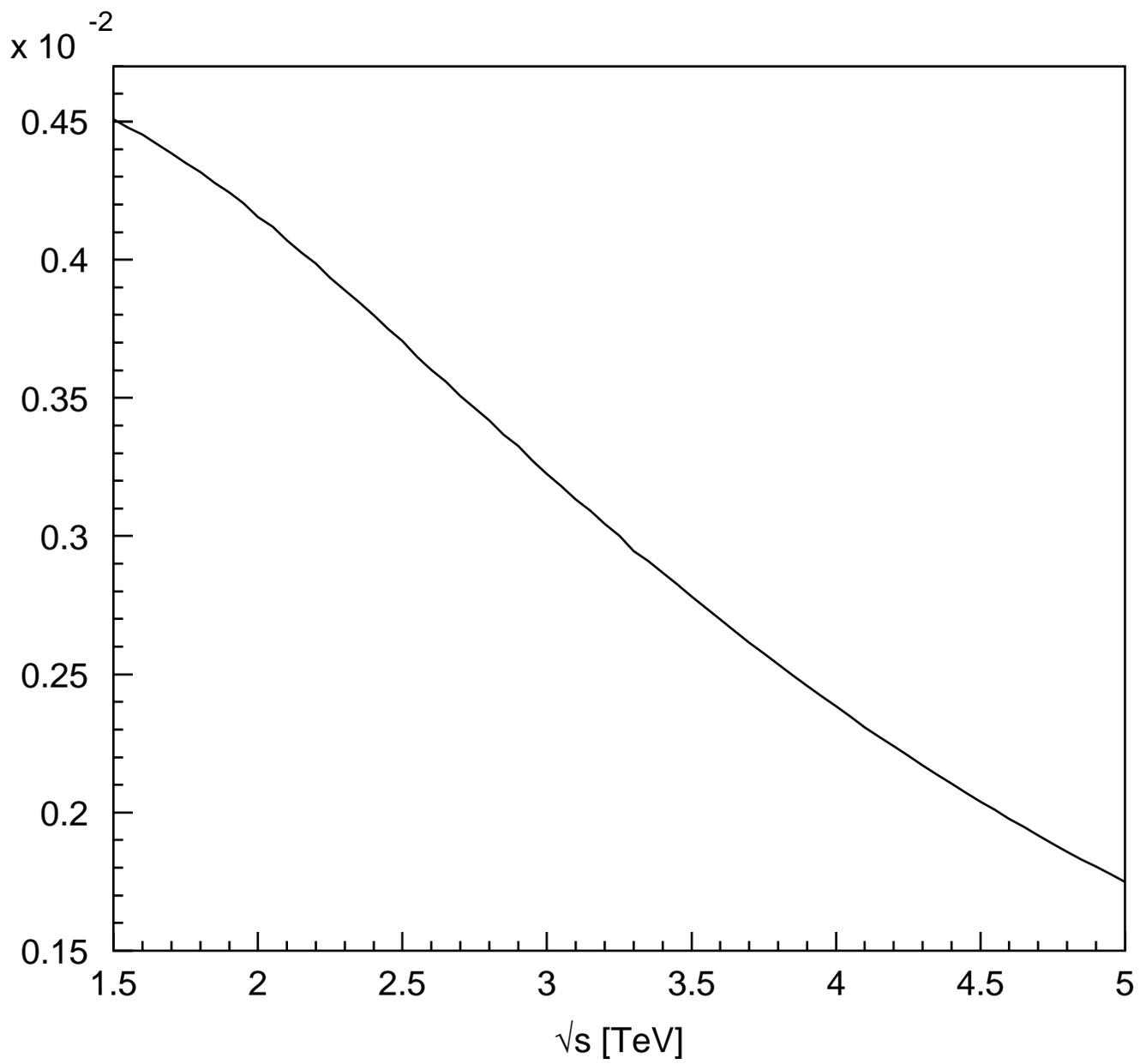
\item[Fig. 5] Correlation ${\cal S}_1$ (defined in Eq. (15)) for
proton-antiproton scattering as a function of the collider c.m. energy
$\sqrt{s}$.
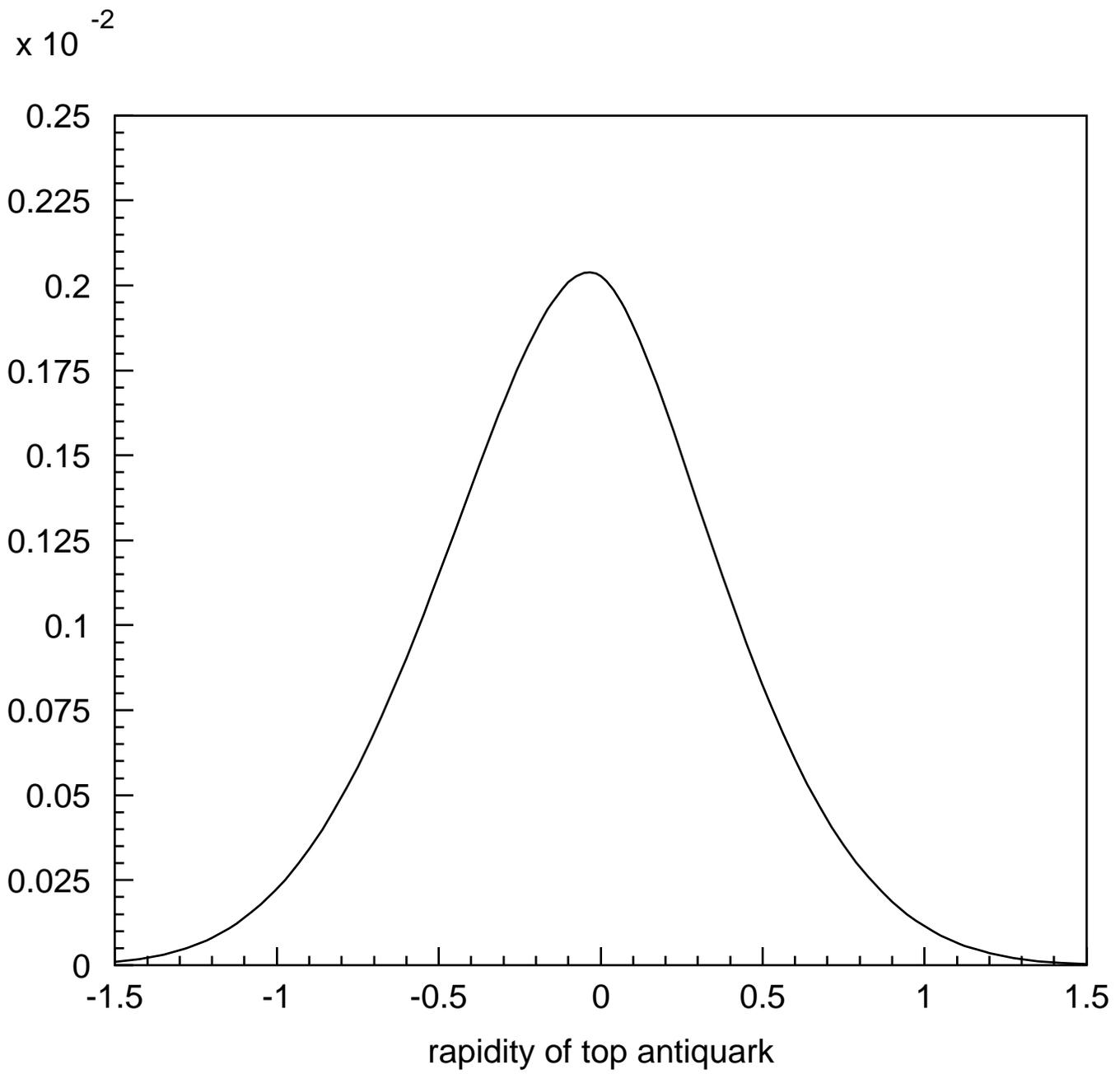
\item[Fig. 6] Rapidity distribution $\langle {\cal O}_1\delta(r_{\bar{t}}
-r_{\bar{t}}')\rangle_{\bar{t}\ell^+}$ for $p\bar{p}$ collisions
at $\sqrt{s}=1.8$ TeV.
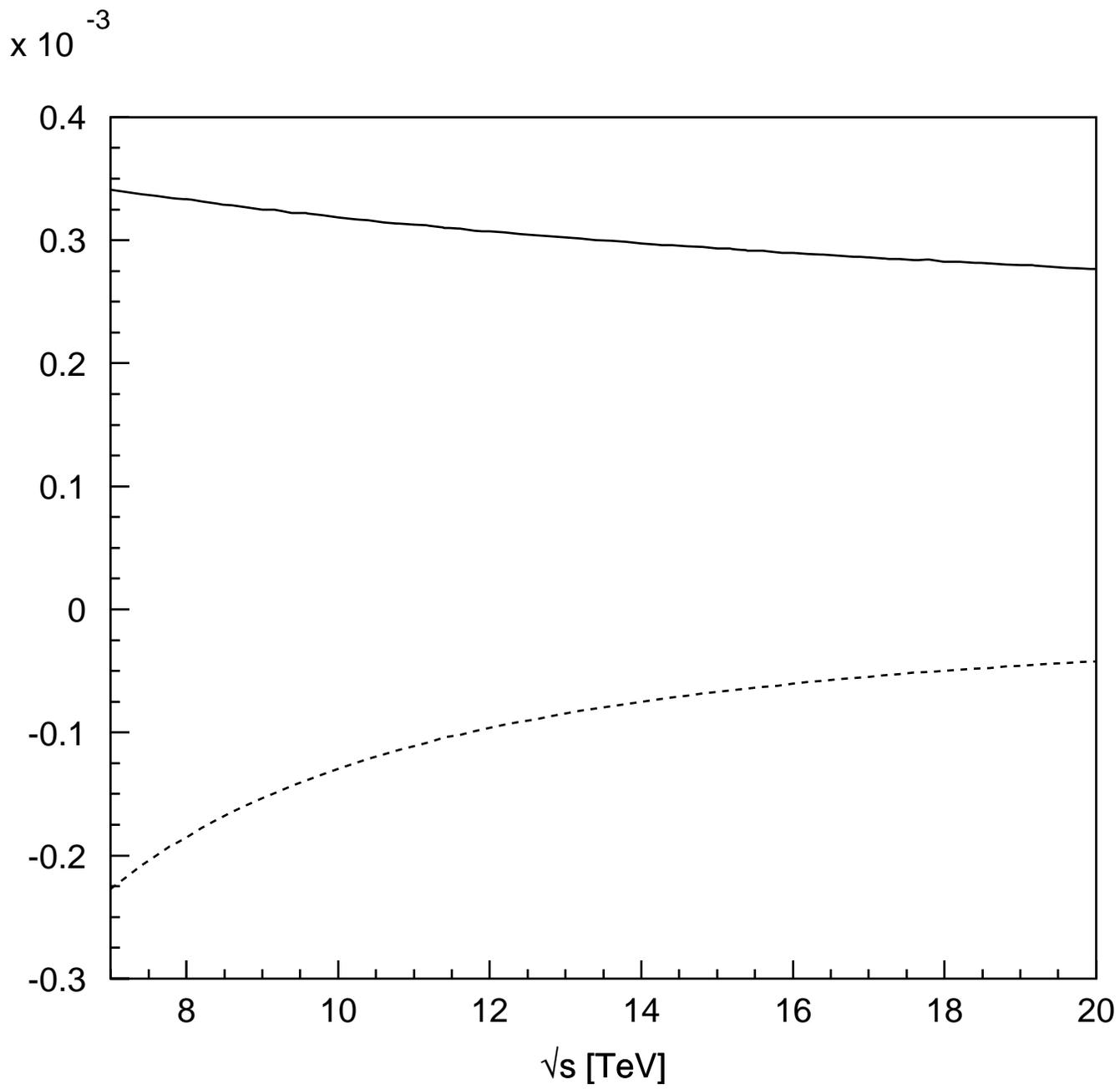
\item[Fig. 7] Correlations ${\cal S}_2$ (full line) and
$\Delta_2$ (dashed line)
(defined in Eqs. (18) and (19), respectively) for proton-proton collisions
as a function of the collider c.m. energy $\sqrt{s}$.
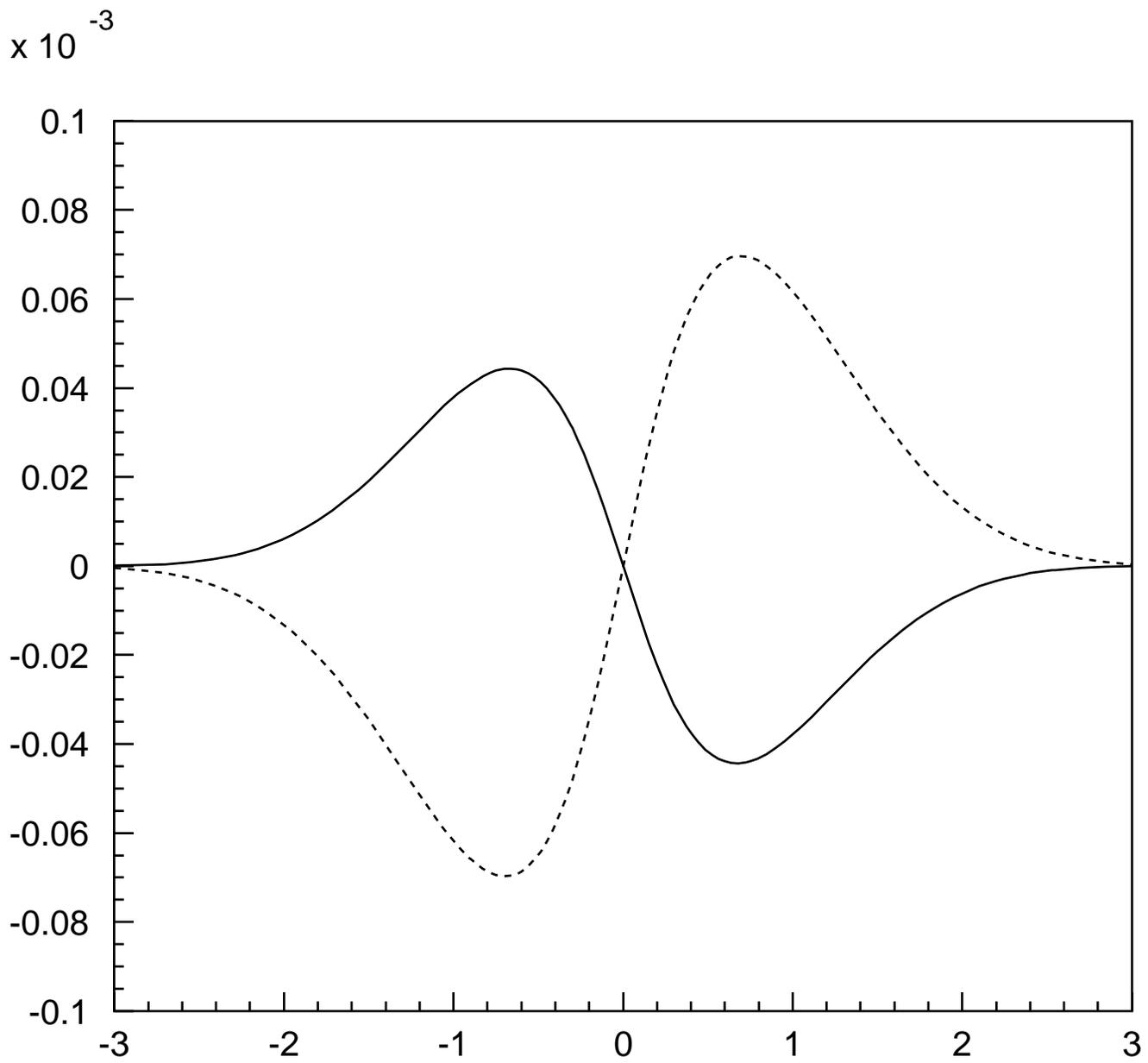
\item[Fig. 8]  Rapidity distributions
$\langle {\cal O}_1\delta(r_{\bar{t}}
-r_{\bar{t}}')\rangle_{\bar{t}\ell^+}$ (full line) and
$\langle \bar{\cal O}_1\delta(r_t
-r_t')\rangle_{t\ell^-}$ (dashed line) for $pp$ collisions
at $\sqrt{s}=14$ TeV.
\end{description}
\end{document}